\documentclass[sigconf]{acmart}
\AtBeginDocument{%
  \providecommand\BibTeX{{%
    \normalfont B\kern-0.5em{\scshape i\kern-0.25em b}\kern-0.8em\TeX}}}

\usepackage{listings}
\usepackage{algorithm}
\usepackage{tikz}
\usepackage[noend]{algpseudocode}
\usepackage{multirow}
\usepackage{hyperref}

\copyrightyear{2021}
\acmYear{2021}
\setcopyright{rightsretained}
\acmConference[SAC '21]{The 36th ACM/SIGAPP Symposium on Applied Computing}{March 22--26, 2021}{Virtual Event, Republic of Korea}
\acmBooktitle{The 36th ACM/SIGAPP Symposium on Applied Computing (SAC '21), March 22--26, 2021, Virtual Event, Republic of Korea}\acmDOI{10.1145/3412841.3442140}
\acmISBN{978-1-4503-8104-8/21/03}

\begin{document}

\title[Discovering Multiple Design Approaches in PA Submissions]{Discovering Multiple Design Approaches in Programming Assignment Submissions}
\author{Nikhila KN}
\orcid{0000-0001-5954-8554}
\affiliation{%
  \institution{International Institute of Information Technology}
  \streetaddress{Electroncis city}
  \city{Bangalore} 
  \state{Karnataka} 
  \country{India}
  \postcode{560100}
}
\email{nikhila.kn@iiitb.org}

\author{Sujit Kumar Chakrabarti}
\affiliation{%
  \institution{International Institute of Information Technology}
  \streetaddress{Electroncis city}
  \city{Bangalore} 
  \state{Karnataka} 
  \country{India}
  \postcode{560100}
}
\email{sujitkc@iiitb.ac.in}

\author{Manish Gupta}
\affiliation{%
  \institution{International Institute of Information Technology}
  \streetaddress{Electroncis city}
  \city{Bangalore} 
  \state{Karnataka} 
  \country{India}
  \postcode{560100}
}
\email{manish.gupta@iiitb.ac.in}


\begin{abstract}
 In this paper, we present a novel approach of automated evaluation of programming assignments~(AEPA) the highlight of which is that it automatically identifies multiple solution
approaches to the programming question from the set of submitted solutions. Our approach does not require the instructor to foresee all the
possible solution approaches and accomplishes this task with little or no
human intervention. This paves the way to multiple fundamental improvements in the way automated evaluation of programming assignments is
done today. We have applied our method on multiple data sets of
practical scale. In all cases, our method was able to detect the solution
approaches employed by the students.
\end{abstract}


\maketitle
\section{Introduction}
    Programming is a skill essential for all professionals, especially in the field of science and technology. In the age of online learning, anyone who is passionate about learning can overcome the challenges and make learning much easier. In online platforms like Massive Open Online Courses(MOOCs), the number of students that can be enrolled in a course is virtually unlimited. Manual methods of evaluation do not scale for such numbers. It is in this context that the need for an automated evaluation tool emerges.

    A majority of the current automated evaluation tools used today are mainly test-case based. The marks assigned to a submitted solution depends on the number of test cases that pass on it. The pass or fail status of a submitted solution depends on a comparison of its output with that of a reference solution provided by the instructor. Following are some of the important drawbacks to this method:
    \begin{enumerate}
    \item \textbf{Over penalisation of silly mistakes.} Novice programmers, who constitute a large section of the students in introductory programming courses, are likely to make silly mistakes, both at the syntax level (e.g. forgetting to close a brace), but also at the semantic level (e.g. forgetting to increment a loop counter leading to infinite loop). While syntax errors can be caught upfront by the compiler, semantic errors may not get caught through preliminary testing. Such silly mistakes are more a result of unfamiliarity or nervousness than of lack of understanding. In practical cases, a human evaluator would always be considerate to such silly mistakes, maybe awarding partial marks. However, testing is very unforgiving to such silly mistakes. Worst still, testing is completely unable to distinguish between silly mistakes and mistakes arising out of lack of conceptual understanding. This makes purely testing based approach fundamentally inappropriate for automated evaluation particularly for entry level programming courses.
	\item \textbf{False negatives.} When a verification method fails to find a defect which actually exists, it is called \emph{false negative}. Testing is well known to be incapable of ensuring the absence of a defect. Likewise, in the context of automated evaluation, it is possible to mark incorrect programs as correct in a system that evaluates assignments only on the basis of test cases. Consider an example where an assignment requires to sort an array using Quick-sort method. The solution can also be implemented by using any other sorting (say, merge-sort) method. The output of this solution would be identical to the that of Quick-sort, even though it is not a correct solution. An evaluation system based on testing would not be able to detect this mistake \footnote{Although, in some cases, there may be ways to design test cases that track the performance of the program and make inferences out of that, these do not yield robust evaluation strategies.}.
    \end{enumerate}
    
 In this paper, we introduce a novel idea for identifying the solution approach followed by a submitted program. Our method has the following unique features:
 \begin{enumerate}
\item Our method identifies a solution approach by mining the set of submitted solutions using node clustering~\cite{Aggarwal2010};
\item Our method practically needs no extra inputs from the instructor to identify novel solution approaches amongst the solutions submitted. It can identify solution approaches which may not even have been foreseen by the instructor.
\item With minimal manual intervention, incorrect solution approaches which would pass the test cases can be identified in a scalable way.
 \end{enumerate}
        
The paper is organised as follows: Section~\ref{sec:me} presents an example to motivate the idea proposed in this paper. Section~\ref{sec:rw} discusses the related work. Section~\ref{s:app} introduces our approach for automated evaluation and briefly explains the architecture and the different steps in the evaluation algorithm. Section~\ref{ppm} explains in detail the method for discovering the multiple design approaches in programming assignments. In Section~\ref{sec:exp} and Section~\ref{s:maxsize}, we present the experimental evaluation and results. Section~\ref{sec:con} concludes the paper with a summary and future work.

\section{Motivating Example}\label{sec:me}

\begin{center}
\begin{tabular}{c}
\begin{lstlisting}[language=c,basicstyle=\ttfamily,keywordstyle=\color{blue}]
int sum(int n) {
  if(n==1)
    return 1;
  else 
    return n+sum(n-1);
 }
\end{lstlisting}
\end{tabular}\\
(a)
\end{center}
\begin{center}
\begin{minipage}{0.3\linewidth}
\centering
\begin{lstlisting}[language=c,basicstyle=\ttfamily,keywordstyle=\color{blue}]
int sum(int limit) {
    int sum=0,k=1;
   for(k=1;k<=limit;k++)
    {
        sum+=k;
    }
    return sum;
 }
\end{lstlisting}

(b)
\end{minipage}%
\hspace{20mm}
\begin{minipage}{0.4\linewidth}
\centering
\begin{lstlisting}[language=c,basicstyle=\ttfamily,keywordstyle=\color{blue}]
int sum(int n) {
    int sum=0,k=1;
    while(k<=n)
    {
        sum+=k;
        k++;
    }
    return sum;
 }
\end{lstlisting}
(c)
\end{minipage}%
\captionof{figure}{Different solution approaches: (a) Recursion; (b), (c): loops}
\label{f:ex}
\end{center}

In this section, we present an example to motivate the work. The same example will be used to illustrate the main parts of our approach in section~\ref{ppm}.

Consider the following programming problem appearing in a fictitious online programming contest: Implement a function \lstinline@sum@ that takes an integer $n$ as input parameter and returns the sum of all natural numbers less than or equal to $n$ using loop, i.e. $1+2+...+n$ .

500 programmers\footnote{all numbers in this example are indicative} participate in the contest. As it turns out, 300 submissions are correct, while the remaining 200 are partially or completely incorrect. Evaluation is done automatically using the following two approaches:

\begin{itemize}
\item Pure testing
\item Our approach
\end{itemize}

Pure testing approach identifies 305 submissions as correct. This includes all correct submission, but also 5 false negatives. These include solutions which have a fundamental conceptual flaw, but output a correct value for all the test cases run. But more importantly, it gives zero marks to 75 incorrect submissions because they fail all the test cases. Examples include submissions which use a while loop to solve the problem, but forget to initialise the counter variable \lstinline@k@ in the beginning, or to increment it inside the body of the loop (as in Figure.~\ref{f:ex}~(c)). In manual evaluation, such solutions would have been awarded partial marks, the mistakes can be categorised as silly mistakes rather than conceptual mistakes.

Our evaluation approach correctly identifies that these 75 incorrect solutions are really attempts to implement a solution which resembles one of the 3 solution approaches as shown in Figure.~\ref{f:ex}. The marks are given, not merely by taking into account the test output, but by considering the structural similarity of each submitted solution to an appropriate approach. These 75 incorrect solutions are not awarded full-marks, as they fail the test cases. However, they are given partial marks which is a more just evaluation.

Importantly, the instructor had provided only one reference solution, say the iterative one in Figure.~\ref{f:ex}~(b). However, our approach automatically detects all the three correct approaches from amongst the 300 solutions. Further, our approach naturally singles out the 5 false negatives whose marks can then be corrected after manual inspection.

This paper focuses on the approach detection module of our evaluation approach, the complete evaluation approach being the topic of a separate paper.

\section{Related Work} \label{sec:rw}
In this section, we discuss the prior art in similarity checking methods and various automated evaluation methods available for evaluating programming assignments and position our work w.r.t. them. 

A majority of the current automated evaluation tools used today are mainly test-case based~\cite{10.1145/2499370.2462195, 10.1145/1163405.1163409, 10.1016/S0360-1315(96)00025-5}. In most of these, the evaluator generates a set of test cases either manually or automatically for the evaluation of the programs. Each test case carries a particular weight of marks and calculates the number of test cases passed and assign grades~\cite{10.1145/1047124.1047427, 10.1145/1163405.1163407, 10.1145/182107.182101}. An alternative of this method is to estimate the average running time when evaluating solutions in test sets. Then it is used to determine the efficiency of each solution and is graded according to those criteria~\cite{ribeiro2009improving}.

Another category of automated grading methods is based on abstract representation of the program like  \emph{abstract syntax tree} (AST), \emph{control flow graph} (CFG), \emph{data flow graph} (DFG)) and \emph{program dependence graph} (PDG) etc. In~\cite{10.1016/j.infsof.2006.03.001} a method of evaluation using semantic similarities is presented. Here,  programs are standardized and then compared to a set of correct program models. They have used \emph{system dependence graphs} as abstract representation and the submissions are evaluated by comparing size, structure, concrete statements and expressions. In~\cite{10.1016/j.compedu.2009.09.005}, a method of evaluation using graph similarity is presented. In this method, after converting submissions to \emph{directed acyclic graph}, the similarity is measured based on the idea that similar nodes will have similar neighbours. The main drawback of the methods based on the similarity check is that they cannot recognise the submissions that deviate from the intended algorithm.  Wang \textit{et al.}~\cite{10.1016/j.infsof.2006.03.001} considered multiple correct reference solutions that were manually created using different approaches to address this problem. The creation of such multiple correct reference solutions using different approach is a tedious task. 

The work presented in the paper~\cite{DBLP:conf/icsoft/ShivamGBB19} introduces another way of automated evaluation using path equivalence checking. The basic idea is to build a Petri net model of the submission and reference solution, and then perform a path equivalence checking on that model for evaluation. 

In addition, we have recently seen the evolution of automated grading using machine learning. The authors of the papers~\cite{srikant, quesindep} have extracted the features that make it possible to understand the functionality of the program from the abstract representation of the program. These features were later used to learn a regression model  to perform grading.

The similarity checking~\cite{10.1016/j.scico.2009.02.007} and plagiarism checking~\cite{10.1145/3313290} has been studied widely for many years and several tools like GPLAG, MOSS, JPLAG, DUP are now available.  These tools are mainly based on different program representations such as Textual, AST, PDG etc.  \emph{Dup}~\cite{514697} is a program which helps to locate duplication or near duplication in source code. Dup uses the textual representation of the program, ignores comments and white spaces and identifies similar sequences of code.  This tool works well for finding the textual similarity.  

In~\cite{Schleimer:2003:WLA:872757.872770, jplag}, the authors have introduced token based similarity checking. These methods tokenize identifiers and keywords from the text representation of the source code. These tokens are then represented as a token sequence and the algorithm that checks the similarity finds similar token sequences and calculates the similarity. These algorithms can detect both textual and lexical similarities and to some extent syntactical / structural similarities.

GPLAG~\cite{GPLAG} is an another tool for plagiarism detection based on program dependence graph. This tool first converts the source code into PDG and examines similar structures of the PDG using subgraph isomorphism. The basic idea of this method is that similar pieces of code have the similar PDG structure. This is capable of detect syntactical/structural similarity. As PDG is the basic structure, this method does not easily scale to multi-procedure programs.

A tool called \emph{SamaTulyata} is presented in the paper~\cite{10.1007/978-3-319-68167-2_8}. The tool works for program equivalence checking based on path equivalence check. It constructs a Petri net based CFGs from the source code and then performs path equivalence check on this CFGs to capture program equivalence.  

The work presented in our paper builds upon several of the above pieces of work. We use testing to segregate the correct solutions from the partially correct/incorrect ones. The output of running test cases is used again for marking. However, much of the approach relies on static similarity estimate. We have used external tools like MOSS and GPLAG to estimate similarity between programs. It is noteworthy that here our objective is to estimate similarity in a positive sense, unlike in the case of referred work which are plagiarism detection tools. We apply feature based similarity estimation in our work both at the similarity estimation stage and marking stage similar to~\cite{srikant}. However, this is done as a step in a larger process. Further, we estimate structural similarity w.r.t. multiple possible reference solutions, rather than a unique one provided by the instructor. Unlike in [21], we automatically detect all correct approaches. This is the most distinctive feature of our contribution. The details of this idea is the major topic of this paper. Unlike in all pure testing or pure static analysis based approaches, we use a combination of testing and static analysis to award the final marks to a solution. This also distinguishes our contribution w.r.t. all evaluation approaches surveyed above.
\section{Automated Evaluation of Programming Assignments} \label{s:app}

\begin{algorithm}
\caption{Automated Evaluation}
\label{a:aepa}
\begin{algorithmic}[1]
\Procedure{evaluate}{$S$}
	\State $G \gets \{s \in S : \forall t \in T\ s(t) \triangleq s_E(t) \}$               \label{pc:G}
	\State $I \gets S \setminus G$                                              \label{pc:I}
	\State $G' \gets$ \Call{sample}{$K$, $G$} \Comment{$r$ = sample size}                                        \label{pc:Gdash}
	\State $\mathcal{G'} \gets$ \Call{make-similarity-graph}{$G'$}
	\State $A \gets$ \Call{cluster}{$\mathcal{G'}$} \Comment{$A=\{C_1, C_2, ..., C_n\}$ s.t. $C_1 \cup C_2 \cup ... \cup C_n = G'$}
	                                                                            \label{pc:A}
	\State $\alpha \gets \bigcup\limits_{c \in C}\{\Call{random-select}{c}\}$   \label{pc:alpha}
	\State $\forall i \in I, S[i] \gets max\bigcup\limits_{c\in\alpha}$\Call{similarity}{$i$, $c$}
	                                                                            \label{pc:S}
	\State $\forall i \in I, M \gets \mathcal{R}(S[i])$                         \label{pc:M}
	\State return $M$
\EndProcedure
\end{algorithmic}
\end{algorithm}

Algorithm~\ref{a:aepa} presents our AEPA approach in the form of pseudo-code. We present the steps of the algorithm in more details below.

\subsection{Gold Standard Detection}  
Let $S$ be the set of all submitted solutions. In this stage, we identify subset $G$ of $S$ such all members of $G$ are completely correct. We call the members of $G$ as \emph{gold standard} solutions.  As shown in line~\ref{pc:G}, $G$ is identified by pushing through each solution $s$ in $S$, a battery of test inputs $t$ which are members of a test input set $T$, and comparing the output of $s$ (given by $s(t)$) with that of the reference solution $s_E$ provided by the evaluator. If $s(t)$ matches $s_E(t)$ for all $t$ in $T$, $s$ is considered to be a completely correct solution and is included in $G$. All gold standard solutions are decided to get full marks.

All the other members of $S$ are identified as incorrect solutions, and are included in the set $I$, as given in line~\ref{pc:I}. Most of the further effort of AEPA is invested on $I$ to determine what partial marks to give each of its members.

As shown in line~\ref{pc:Gdash} of algorithm~\ref{a:aepa}, next, we sample a subset of $G$. This gives us $G'$. This is to contain the computational cost of constructing the similarity graph which is an expensive step. We present further details on this aspect in Section~\ref{s:maxsize}.

\subsection{Approach Detection}
As summarised in line~\ref{pc:A}, to detect the different approaches present in $G'$, we use node clustering. As a first step, we compute a complete graph $\mathcal{G'}$, in which each node represents a member of $G'$ and the weight of each edge between nodes $p_i$ and $p_j$ in $\mathcal{G'}$ given by $w_{ij}$ is the similarity between the the two nodes (solutions). For every pair of solutions $p_i$, $p_j$ in $G'$, we compute the similarity between the two, using a method termed as \textsc{similarity} in the algorithm. In our current implementation, we use an off-the-shelf program similarity detection tool for \textsc{similarity}. This gives us a complete graph $\mathcal{G}_{G'}$. $\mathcal{G}_{G'}$ is given as the input to a graph clustering algorithm. This clusters the graph giving a set $A$. Each element of $A$ is a set of nodes of $G'$ that represents one solution approach. Most of the rest of this paper focuses on this step of our automated evaluation approach.

From each set in $A$, we pick one member at random as the representative of the approach of that set to create the set $\alpha$ of approach representatives, as shown in line~\ref{pc:alpha}.

\subsection{Evaluation of Incorrect Solutions}
To identify the approach of the incorrect solutions in set $I$, we do the following with each member $i$ of $I$: using \textsc{similarity}, we compute the similarity of $i$ with each member of $\alpha$. The approach represented by $\alpha_j$, the member of $\alpha$ whose similarity with $i$ is the highest among all members of $\alpha$, is considered to be the approach followed by $i$. The overall similarity measure of $i$, given as $S[i]$, is the similarity of $i$ with $\alpha_j$.

The intuition behind the above step is that the approach attempted in the solution $i$ would very likely match one of the approaches in $\alpha$, say $\alpha_j$. Also, the similarity of $i$ with $\alpha_j$ is also very likely to be higher than that with any other member of $\alpha$.

\subsection{Assigning Marks}
As shown in line~\ref{pc:M}, we use a pre-trained regression model $\mathcal{R}$ to compute the marks of each member $i$ of $I$ as $M[i] = \mathcal{R}[i]$. $M$ is returned as the final result. Final marks are calculated as a combined outcome of weighted combination of both the structural similarity of the submitted solution with the respective gold standard solution, and its performance with the test cases. The details of this module are outside the scope of this paper.

In this paper, we focus our attention to the problem of approach detection. Details of this method are presented in section~\ref{ppm}. The other aspects of our automated evaluation approach are subjects of separate papers.

\section{Approach Detection Method}\label{ppm}
When given a programming question, each student may solve it by using different approaches. By \emph{approach}, we mean the specific design or implementation strategy used by the student. Here, we propose a novel method to identify the approaches from a set of submitted solutions. The basic idea is to construct a  complete graph called similarity graph and perform node clustering on it. Nodes in the graphs represent the solutions in the submitted set and edges represent the structural similarity between each solutions.

\begin{figure}[ht!]
     \centering
     \resizebox{\columnwidth}{!}{
     \begin{tikzpicture}[correct/.style={circle, draw=black, minimum size=3mm, text=black}, clusters/.style={circle, draw=black, minimum size=3mm, text=black,text width=4.5mm}, module/.style={rectangle, rounded corners, minimum width=1cm, minimum height=1.25cm,text centered, draw=black,  text=black,text centered, text width=2.5cm},gnode/.style={circle, draw=black, minimum size=3mm, text=black},scale=0.8, every node/.style={scale=0.8}]  
            \node [correct] (s1) at (0,1) {$G_{S_1}$};
	    \node [correct] (s2) at (0,-0.25) {$G_{S_2}$};
	    \node [] (dots) at (0,-1){$\vdots$};
	    \node [correct] (s3) at (0,-2) {$G_{S_n}$};
	    \node [module]  (s5) at (2.5,-.5) {Similarity Check};
	    \node [module]  (s6) at (5.75,-.5) {Graph Generator};
	    \node [module]  (s10) at (9,-.5) {Node Clustering};
	    \draw[black, thick, ->] (s2) -- (s5);
	    \draw[black, thick, ->] (s1) -- (s5);
	    \draw[black, thick, ->] (s3) -- (s5);
            \node [clusters] (s11) at (11.5,1) {$C_1$};
	    \node [clusters] (s12) at (11.5,-0.25) {$C_2$};
	    \node [] (dots) at (11.5,-1){$\vdots$};
	    \node [clusters] (s13) at (11.5,-2) {$C_k$};
	    \draw[black, thick, ->] (s10) -- (s11);
	    \draw[black, thick, ->] (s10) -- (s12);
	    \draw[black, thick, ->] (s10) -- (s13);
 	    \draw[black, thick, ->] (s5) -- (s6);
            \draw[black, thick, ->] (s6) -- (s10);
	\end{tikzpicture}}
    \caption{Approach Detection Framework}
     \label{fig:workflow}
 \end{figure}
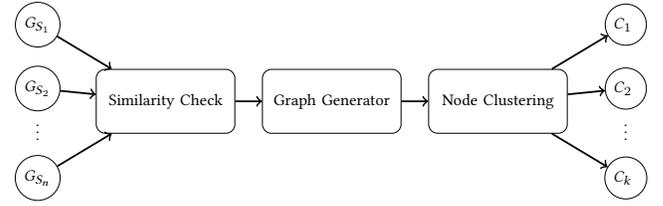

Our hypothesis is that node clustering on the same graph leads to similar approaches to cluster together. To validate this hypothesis, we took data from  ideal clustering (manual clustering) which we consider as ground truth and measured the effectiveness between the clusters formed using our method and the ground truth. The framework of our approach detection method is shown in Figure.~\ref{fig:workflow}.

\subsection{Similarity Graph}\label{s:sc}

This step detects the structural/syntactic similarity between programs. This similarity is later used for clustering. To understand more about structural similarity, consider the example programs shown in Figure.~\ref{f:ex}. These example programs are meant to find the sum of first $n$ natural numbers, i.e. they are functionally similar. The program in Figure.~\ref{f:ex}~(a) uses recursion and the two programs in Figure.~\ref{f:ex}~(b) and (c) use iterative structure using loops. Therefore, the two programs in Figure.~\ref{f:ex}~(b) and (c) structurally similar to each other but are different from the program in Figure.~\ref{f:ex}~(a).

We use an off-the-shelf program similarity estimation method to compute the \emph{similarity graph} ($\mathcal{G}$). The similarity estimation technique is modelled as a function $\textsc{similarity} : program \times program \rightarrow [0, 1]$ which assigns a similarity value between 0 and 1 to two programs given as inputs to it. A similarity value 1 means that the two programs are identical and a similarity value 0 means that the two programs are completely different. $\mathcal{G}$ is a complete undirected weighted graph in which each node corresponds to a correct solution. Each edge $e_{ij}$ in $\mathcal{G}$ between nodes $n_i$ and $n_j$ is labelled with the similarity value $\textsc{similarity}(P_i, P_j)$ where $P_i$ and $P_j$ are programs corresponding to nodes $n_i$ and $n_j$ respectively.


\emph{Example:} Our example data set \emph{First-n-Sum-Sample} mentioned in the section~\ref{sec:me} contains over 500 submissions. Among these 305 are correct submissions identified by the test-case method. From these correct solutions, we have to find out all the approaches that students have taken to solve the problem. For this purpose, we have created a similarity graph which is a complete weighted undirected graph $\mathcal{G}(V,E)$. 
Here, $|V|= 305$ and $|E|= 46360$.
The edges $(n_i, n_j) \in E$ are the pairwise combinations of each node and the edge weight represent the  $\textsc{similarity} (P_i,P_j) \in [0,1]$ where $(P_i,P_j)$ is the submissions programs corresponding to the nodes $n_i,n_j \in V$.
\subsection{Node Clustering} \label{s:nc}
In this module we cluster solutions that have high degree of similarity together.

\begin{definition}{Node Clustering:}
         It is a method of determining the dense areas of the graph based on the edge behavior. Edge behaviour can be either a distance value or a similarity value.
        \end{definition}
        
        \begin{definition}{Gold standard solution:}
        A solution which passes set of all test cases are considered to be a completely correct solution and it is known as the gold standard solution.
        \end{definition}
        
        \begin{definition}{Gold standard solution set(G):~}
        Let $S$ be a set of all submitted solutions, $G \subseteq S$ such that all members in $G$ are the gold standard solutions.
        \end{definition}

We use an off-the-shelf clustering algorithm to perform node clustering of the similarity graph $\mathcal{G}$. We cluster $\mathcal{G}$ using weighted node clustering. Each resultant cluster will have nodes where the similarity between them are high. Since the nodes represent the gold standard solutions and the edges represent the degree of similarity, each cluster contains very similar gold standard solutions. Our hypothesis states that highly similar solutions have common approaches, so it is highly probable that each cluster represents a unique approach that was taken by the student to solve the problem. To detect clusters, we tried Louvain modularity~\cite{louvian}, IPCA~\cite{ipca}, and spectral clustering~\cite{10.1007/s11222-007-9033-z}. 

\section{Experimental Evaluation}\label{sec:exp}
In the current set up, we use off-the-shelf tools for both similarity estimation and clustering. Further, we carry out these activities for discovering the ground truth data. For similarity estimation, we used MOSS~\cite{Schleimer:2003:WLA:872757.872770} and GPLAG~\cite{GPLAG}. For clustering, we use Louvain modularity~\cite{louvian}, IPCA~\cite{ipca}, and spectral clustering~\cite{10.1007/s11222-007-9033-z}. During our experiments, we observed differences in the performance of each of these methods over the different programs in our data set. It was also evident that different combinations of similarity estimation tools and clustering tools yielded different results. While we did not attempt to try all combinations exhaustively, we were interested in setting up experiments which could help evaluate the various tools used and their combinations. Therefore, we designed our experiments with the objective of addressing the following research questions:
\begin{enumerate}
    \item $RQ_1$: Which similarity checking algorithm is well suited for discovering the solution approaches?\label{ex:q1}
    \item $RQ_2$: Which clustering algorithm is acceptable to obtain exact solution approaches?\label{ex:q2}
	\item $RQ_3$: How effective is our method in identifying solution approaches?\label{ex:q3}
\end{enumerate}

    \subsection{Data}
We intend to use our evaluation method primarily for fresher level programming tests. Hence, for our experiments, we prepared a data set with the following properties:
        \begin{enumerate}
        \item short length ($\le$ 50 lines)
        \item Basic programming language~(PL) features are used e.g. integer values as input/output, conditional statements, loops etc are present
        \item Advanced PL features are excluded, e.g. function definitions/calls, structures, pointer manipulations etc. 
        \end{enumerate}

         We used three data sets from competitive coding websites such as HackerRank~\cite{hackerrank} and Codeforces~\cite{CodeForces} which offer programming courses. These are summarised in Table~\ref{tab:data set}. 
         Also, ground truth is needed to measure the quality of similarity estimation and the clustering. For this, the similarity is calculated and clustering is performed through a manual method presented below.

\begin{table}[htp]
\caption{Feature rubric for manual similarity estimation for Program shown in Figure.~\ref{f:ex}~(b)}
\begin{tabular}{ c  c }
\toprule
Feature & Feature weight \\
\midrule
Variable declaration & 0.1 \\
Loop matches & 0.1\\
Loop condition matches & 0.2 \\
Calculation inside loop & 0.2\\
Sum calculation & 0.3\\
Output Statement & 0.1 \\
Each extra statements & -0.025\\
\bottomrule
\end{tabular}
\label{t:rubric}
\end{table}

 \begin{itemize}
         \item \textbf{Similarity estimation:} For each of the programs in our data set, we collaboratively designed a feature rubric. As an example, the feature rubric for Figure.~\ref{f:ex}~(b) is presented in Table~\ref{t:rubric}. If $p_1$ and $p_2$ (the two gold standard programs whose similarity is being estimated) matched on a particular feature, the similarity score goes up by the feature weight in the rubric. The total similarity score is the sum of all the feature scores. Two independent experts (helpful research scholars outside the team) were provided with the data set and the rubric. They independently estimated the similarity between each pair of programs in the data set. As a final step, they tallied the scores they provided to each pair.  For data points in which the estimated similarity was less than or equal to 0.2, the average of the similarity scores assigned by each expert was taken as the final similarity score. Where the discrepancy between the individual estimated similarities exceeded 0.2, they got together and reached a consensus through discussion. This was needed for 10-15\% cases in each data set.
         \item \textbf{Clustering:} A group inspection of the submitted programs was done to identify the solution approach. No formalised process was needed here. Various approaches were identified through manual inspection. The manually estimated similarity scores were passed through automated clustering (Louvain). Interestingly, the clustering thus computed matched the manually identified approaches exactly for all data sets.
         \end{itemize}

        \begin{table}[htp]
            \centering
            \caption{Data sets}
            \label{tab:dataset}
            \resizebox{\columnwidth}{!}{
            \begin{tabular}{cccc}
               \toprule
                 \textbf{Data set}& \textbf{Size} & \textbf{\#Correct Solutions} & \textbf{\#Expected Clusters} \\ 
                \midrule
                  Let's Watch Football& 44 & 27 & 4 \\
		          First-n-Sum-Sample & 500 & 305 & 3 \\
		          First-n-Sum-Original & 1415 & 1220 & 3 \\
                  Recursion-in-c & 2280 & 1280 & 5 \\                  
		          For-loop-in-c & 2480 & 1280 & 4\\
               \bottomrule
            \end{tabular}}
        \end{table}

        \subsection{Experiments}
        \subsubsection{Experiment 1: }
        Here, we would like to address $RQ_{\ref{ex:q1}}$. We used methods MOSS and  $\gamma$-isomorphism~\cite{GPLAG} to estimate the similarity. To measure the effectiveness of these methods, we compare the methods with an ideal similarity checker~(a manual method) and the results are shown in Table~\ref{tab:stat_values}. From  Table~\ref{tab:stat_values}, we can see that the $\gamma$-Isomorphism gives the best correlation and the least error are made for the first three problems and MOSS works well for the remaining data set. This is because MOSS works better for large size codes. The programs in the first three data sets are small~(LOC $\le$15). One of our current efforts is to come out with structural similarity estimation algorithms to improve upon the results obtained from off-the-shelf methods like $\gamma$-isomorphism.

         \begin{table}[htp]
            
            \caption{Comparison of MOSS and $\gamma$-isomorphism with Ideal method}
            \label{tab:stat_values}
            \resizebox{\columnwidth}{!}{
            \centering
            \begin{tabular}{ccccc}
               \toprule
                 \textbf{Data set} & \textbf{Method} & \textbf{MAE}&\textbf{EV}&\textbf{SC}\\ 
                \midrule
                  \multirow{2}{*}{Let's Watch Football}& $\gamma-$Iso & 0.3216 & -0.1193 & -0.088\\ 
                  & MOSS & 0.3159 & -0.4883 & 0.140\\
                  \multirow{2}{*}{ First-n-Sum-Sample} & $\gamma-$Iso& 0.2213 & 0.2234 & 0.763 \\
                  & MOSS & 0.5643 & -0.0273  & 0.293 \\
                  \multirow{2}{*}{First-n-Sum-Original} & $\gamma-$Iso& 0.2144 & 0.2354 &0.798 \\
                  & MOSS & 0.5556 & -0.0187  & 0.338 \\
		  \multirow{2}{*}{Recursion-in-c}& $\gamma-$Iso & 0.4178 & 0.02822 & 0.246 \\
		  & MOSS & 0.1173 & 0.8221  & 0.892\\
		  \multirow{2}{*}{For-loop-in-c}& $\gamma-$Iso & 0.7149&-3.4166 & 0.198\\
		  &MOSS & 0.0771 & 0.7881 & 0.724\\
         	  
               \bottomrule
            \end{tabular}}
            \begin{tabular}{c}
                 MAE: Mean Absolute Error, EV: Explanatory Variance, \\ SC: Spearman's Correlation
            \end{tabular}
        \end{table}
        
        \subsubsection{Experiment 2:}
        In this section we address $RQ_{\ref{ex:q2}}$ and $RQ_{\ref{ex:q3}}$. 
        Here, we are looking for clusters that do not have cluster overlapping. If cluster overlapping is formed, this means that the same solution appeared in multiple approaches. This is not permissible in our method. In addition, in practice, it would be impossible to say how many approaches exist before clustering. Therefore, we tried Louvain and IPCA  and exclude the methods like \emph{spectral clustering}~\cite{10.1007/s11222-007-9033-z} which use the number of clusters as input.
 
\begin{center}
\begin{minipage}{0.45\linewidth}
\centering
    \includegraphics[width=\linewidth]{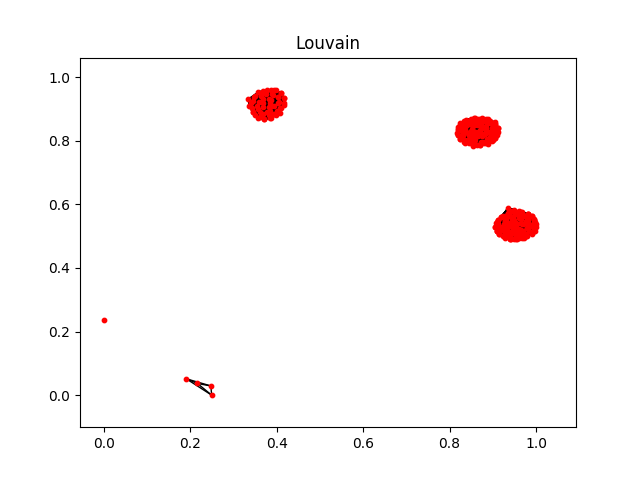}
    (a)
\end{minipage}%
\hspace{5mm}
\begin{minipage}{0.45\linewidth}
\centering
            \includegraphics[width=\linewidth]{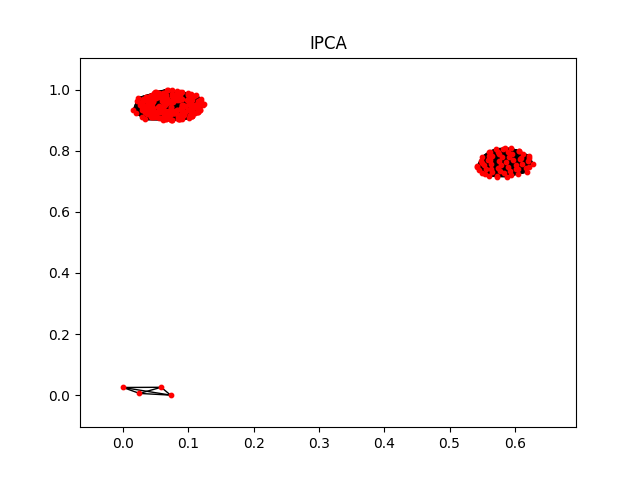}

(b)
\end{minipage}%
\captionof{figure}{Clustering output of Data set First-n-Sum-Sample: (a) Louvain (b) IPCA}
\label{f:co}
\end{center}

\begin{table}[htp]
        \caption{Comparison of Clustering Output}
    \label{tab:result}
        \begin{center}
\resizebox{\columnwidth}{!}{
\begin{tabular}{c c c c c c}
\toprule
\textbf{Data set}& \textbf{\#GS}& \textbf{\#EC} & \textbf{SCM}& \textbf{CM} & \textbf{\#AC}\\
\midrule
\multirow{4}{*}{Let's Watch Football} & \multirow{4}{*}{27} &\multirow{4}{*}{4} & \multirow{2}{*}{MOSS} & Louvain & 23 \\
    &  &   & & IPCA&4\\%
    \cline{4-6}
    &  & & \multirow{2}{*}{Gamma} & Louvain & 5 \\
    & &  & & IPCA&6 \\
    \hline
     \multirow{4}{*}{First-n-Sum-Sample} & \multirow{4}{*}{305} &\multirow{4}{*}{3} & \multirow{2}{*}{MOSS} & Louvain &  12\\
    & &  & & IPCA& 3\\
    \cline{4-6}
    &  & & \multirow{2}{*}{Gamma} & Louvain & 5 \\
    & &  & & IPCA& 4\\
    \hline
    \multirow{4}{*}{First-n-Sum-Original} & \multirow{4}{*}{1220} &\multirow{4}{*}{3} & \multirow{2}{*}{MOSS} & Louvain &  12\\
    & &  & & IPCA& 3\\
    \cline{4-6}
    &  & & \multirow{2}{*}{Gamma} & Louvain & 5 \\
    & &  & & IPCA& 4\\
    \hline
    \multirow{4}{*}{Recursion in C} & \multirow{4}{*}{1280} &\multirow{4}{*}{5} & \multirow{2}{*}{MOSS} & Louvain & 4 \\

    &  &   & & IPCA& 3\\%
    \cline{4-6}
    &  & & \multirow{2}{*}{Gamma} & Louvain & 2\\
    & &  & & IPCA& 1\\
    \hline
    \multirow{4}{*}{For loop in C} & \multirow{4}{*}{1280} &\multirow{4}{*}{4} & \multirow{2}{*}{MOSS} & Louvain & 5 \\
    &  &   & & IPCA& 8\\%
    \cline{4-6}
    &  & & \multirow{2}{*}{Gamma} & Louvain & 2\\
    & &  & & IPCA& 2\\
    \bottomrule
\end{tabular}}
\begin{tabular}{c}
\#GS: Size of Gold standard solution set, \#EC: Expected Clusters,\\ SCM: Similarity checking method, CM: Clustering method \\ \#AC: Actual Cluster
\end{tabular}
\end{center}
\end{table}
        
        Five experiments were performed using five different size data sets and the results are given in Table~\ref{tab:result}.
        
        We observe that the number of clusters formed by IPCA was close to the expected clusters.
        Also, we found cluster overlapping on clusters formed by IPCA.  Consider the Figures.~\ref{f:co}~(a) and \ref{f:co}~(b) which show the clusters formed by the Louvain and IPCA respectively. The input graph weight is calculated by $\gamma$-Iso method. The small cluster in both figures represents the ``false negatives", while the larger cluster in Figure.~\ref{f:co}~(b) is actually two overlapped clusters. The clusters formed by Louvain do not overlap with each other, and each cluster represents a unique approach. In addition, the accuracy of clustering can be improved by improving the accuracy of the similarity check. Based on these observations, we noticed that the Louvain clustering consistently captures unique clusters, which are really different approaches.

\section{Maximum Size of Similarity Graph} \label{s:maxsize}
The cost of preparing the similarity graph is quadratic in the number of gold standard solutions. For $n$ gold standard solutions,  we need $\binom{n}{2}$ computations of pairwise similarity. This may lead this step to become unscalable when the number of gold standard solutions exceeds a certain value. For example, in our observation, for a data set with around 1000 gold standard solutions, the time required to compute the similarity graph may extend to several weeks on a normal PC, which may be unaffordable in most cases. 

However, we have found that it is not necessary to use all the gold standard solutions for computing the similarity graph. If the number of gold standard solutions exceeds $K = 720$, our experiments have revealed that it suffices to randomly select $K$ gold standards solutions out of that and construct the similarity graph from that set.

Having an experimentally validated value of $K$ is critical to the scalability of our approach. With the assurance of having $K$, we are guaranteed that we need use no more than $K$ nodes to prepare our similarity graph. Thus, similarity graph construction despite being a $O(N^2)$ step does not become a hurdle in our approach scaling to classes of arbitrary size.

In the rest of the section, we present our experimental approach to estimate $K$.

\subsection{Inflating the Data}
All output of our algorithms are compared against the ground truth which is manually estimated values, e.g. similarity between programs. Due to resource constraints, it was not feasible for us to work with gold standard solution sets of large sizes. For example, for a gold standard solution set of size 1000, we would need to estimate similarity between $\binom{1000}{2} = 499500$ pairs of solutions. With approximately half a minute needed to process each pair manually, this would take 249750 minutes -- corresponding to nearly two human years of work!

To get around this problem, we took a gold standard solution set with $\approx 300$ solutions, computed the similarity graph $\mathcal{G}<V, E>$ for the same, and inflated this graph to three times its size using repeated \emph{graph join}~\cite{diestel2006graph} of $\mathcal{G}$ with itself. The resultant graph $\mathcal{G}'<V', E'>$ is also a complete graph such that $|V'| = 3|V|$. There are three nodes in $V'$ corresponding to every node in $V$. For any node $v' \in V'$, we denote its counterpart in $v \in V$ as $M_V(v')$. Likewise, each edge $e' \in E'$ has a counterpart $e \in G$ denoted by $M_E(v')$. The weights of the edges in $E'$, with its source and destination nodes denoted by $e'.s$ and $e'.d$ respectively, are computed as follows:

\begin{equation*}
W(e') =
\begin{cases}
W(M_E(e')) & \text{if } M_V(e'.s) \neq M_V(e'.d),\\
1 & \text{otherwise}\\
\end{cases}
\end{equation*}  

\subsubsection{Justification}
We believe that the method of inflating the similarity graph size presented above is useful and justified. Given that the programs we are dealing with are small, a similarity graph created from an originally large set of gold standard solution tends to have very similar information as another similar sized graph generated from a smaller sized data using repeated graph joins. Our observation on all large data sets we have seen so far also indicates without exception that we cease to see any noticeable variations in solutions after having processed a certain number of them.

\subsection{Experimental estimation of the Maximum Size}

\begin{definition}{Clustering Agreement:}
Let $C$ and $C'$ be a clustering defined on a graph $\mathcal{G}$ and $\mathcal{G}'$ respectively, where $\mathcal{G}'$ is an induced subgraph of $\mathcal{G}$. The clustering $C'$ is said to \emph{agree} with $C$ (denoted by $C' \sqsubseteq C$) iff:
\begin{itemize}
\item The number of clusters in $C'$ is equal to the number of clusters in $C$.
\item For each node cluster $c_i$ in $C$, there exists a node cluster $c'_i$ in $C'$ such that $c'_i \subseteq c_i$.  
\end{itemize}
\end{definition}

With the inflated similarity graph $\mathcal{G}$ we first compute the cluster, denoted by $C$. We randomly sample a subgraph $\mathcal{G}_i$ of $\mathcal{G}$ having $k_i$ nodes. 
If, for 5 repetitions of this step, we find $\mathcal{G}_i$ to result clusters $C'_i$ such that $C'_i \sqsubseteq C$, then we conclude that $k_i \ge K$, the maximum size required for the similarity graph. We repeat this step for progressively smaller values of $k_i$. The smallest $k_i$ for which $C'_i \sqsubseteq C$ is considered as $K$.

Note that $K$ is an experimental estimate of the maximum required size of the similarity graph for programs which are comparable to the ones we have considered. $K$ may assume other values for programs which are much larger, or use a different programming language. The actual relation between $K$ and the characteristics of the program in consideration is an interesting research question which we have not tried to address directly here. However, it is noteworthy that the approximate range within which we have experimentally located the value of $K$ is precise enough to be practically useful for our purpose.

\begin{table}[htp]
\caption{Experimental estimation of maximum size of similarity graph}
\label{t:maxsize}
\begin{tabular}{ c  c  c c c}
\toprule
Example & Number of nodes & $K$ & \#EC & \#AC \\
\midrule
First-n-Sum-Original & 1220  & 720 & 3 & 5 \\
Recursion in C & 1280 & 700 & 5 & 4 \\
For loop in C & 1280 & 690 & 4 & 5 \\
\bottomrule
\end{tabular}
\begin{tabular}{c}
\#EC: Expected Clusters, \#AC: Actual Clusters
\end{tabular}
\end{table}
We conducted this experiment with different data sets as presented in Table~\ref{t:maxsize}. We observe, that for all data sets, $K\approx700$ for the programs we have considered.

\section{Conclusion}\label{sec:con}
Neither a purely testing based approach, nor a purely static analysis based approach can completely mimic the judgements of a human evaluator for evaluating programming assignments. In this paper, we have introduced a novel approach for automated evaluation of programming assignments that combines testing and static analysis approach including machine learning. One of the most important features of this approach is its ability to discover the various solution approaches to the programming problem done by mining the set of submitted solutions. This frees the instructor/evaluator from the unrealistic requirement of imagining \emph{a priori} all possible approaches that may be employed by the students to solve a programming problem. This paper presents in detail this aspect of our automated evaluation approach. Our experiments establish that our method successfully discovers the different solution approaches in the set of submitted solutions. We have presented multiple experiments to provide pragmatic pointers that may help in tuning this step for increased reliability (Section~\ref{sec:exp}). We have also presented an experimental evidence that the computational complexity of our approach detection method is effectively independent of the number of submitted solutions (Section~\ref{s:maxsize}). This means that our method scales well to arbitrarily large class sizes.

The approach detection method presented in the paper is a part of an automated evaluation approach introduced in Section~\ref{s:app} of this paper. The approach essentially has three main modules: \emph{similarity estimation}, \emph{approach detection} and \emph{marking}. Our current research efforts are directed towards developing algorithms for each of these modules. Concurrently, design of the overall approach is about combining each of these three modules in the best configuration. This will be one of the focus areas of our future efforts in this project.

\bibliographystyle{ACM-Reference-Format}
\bibliography{ref} 


\begin{thebibliography}{26}


\ifx \showCODEN    \undefined \def \showCODEN     #1{\unskip}     \fi
\ifx \showDOI      \undefined \def \showDOI       #1{#1}\fi
\ifx \showISBNx    \undefined \def \showISBNx     #1{\unskip}     \fi
\ifx \showISBNxiii \undefined \def \showISBNxiii  #1{\unskip}     \fi
\ifx \showISSN     \undefined \def \showISSN      #1{\unskip}     \fi
\ifx \showLCCN     \undefined \def \showLCCN      #1{\unskip}     \fi
\ifx \shownote     \undefined \def \shownote      #1{#1}          \fi
\ifx \showarticletitle \undefined \def \showarticletitle #1{#1}   \fi
\ifx \showURL      \undefined \def \showURL       {\relax}        \fi
\providecommand\bibfield[2]{#2}
\providecommand\bibinfo[2]{#2}
\providecommand\natexlab[1]{#1}
\providecommand\showeprint[2][]{arXiv:#2}

\bibitem[\protect\citeauthoryear{Aggarwal and Wang}{Aggarwal and Wang}{2010}]%
        {Aggarwal2010}
\bibfield{author}{\bibinfo{person}{Charu~C. Aggarwal} {and}
  \bibinfo{person}{Haixun Wang}.} \bibinfo{year}{2010}\natexlab{}.
\newblock \bibinfo{booktitle}{\emph{A Survey of Clustering Algorithms for Graph
  Data}}.
\newblock \bibinfo{publisher}{Springer US}, \bibinfo{address}{Boston, MA},
  \bibinfo{pages}{275--301}.
\newblock
\urldef\tempurl%
\url{https://doi.org/10.1007/978-1-4419-6045-0_9}
\showDOI{\tempurl}


\bibitem[\protect\citeauthoryear{{Baker}}{{Baker}}{1995}]%
        {514697}
\bibfield{author}{\bibinfo{person}{B.~S. {Baker}}.}
  \bibinfo{year}{1995}\natexlab{}.
\newblock \showarticletitle{On finding duplication and near-duplication in
  large software systems}. In \bibinfo{booktitle}{\emph{Proceedings of 2nd
  Working Conference on Reverse Engineering}}. \bibinfo{pages}{86--95}.
\newblock


\bibitem[\protect\citeauthoryear{Bandyopadhyay, Sarkar, Sarkar, and
  Mandal}{Bandyopadhyay et~al\mbox{.}}{2017}]%
        {10.1007/978-3-319-68167-2_8}
\bibfield{author}{\bibinfo{person}{Soumyadip Bandyopadhyay},
  \bibinfo{person}{Santonu Sarkar}, \bibinfo{person}{Dipankar Sarkar}, {and}
  \bibinfo{person}{Chittaranjan Mandal}.} \bibinfo{year}{2017}\natexlab{}.
\newblock \showarticletitle{SamaTulyata: An Efficient Path Based Equivalence
  Checking Tool}. In \bibinfo{booktitle}{\emph{Automated Technology for
  Verification and Analysis}}, \bibfield{editor}{\bibinfo{person}{Deepak
  D'Souza} {and} \bibinfo{person}{K.~Narayan~Kumar}} (Eds.).
  \bibinfo{publisher}{Springer International Publishing},
  \bibinfo{address}{Cham}, \bibinfo{pages}{109--116}.
\newblock
\showISBNx{978-3-319-68167-2}


\bibitem[\protect\citeauthoryear{Blondel, Guillaume, Lambiotte, and
  Lefebvre}{Blondel et~al\mbox{.}}{2008}]%
        {louvian}
\bibfield{author}{\bibinfo{person}{Vincent~D Blondel},
  \bibinfo{person}{Jean-Loup Guillaume}, \bibinfo{person}{Renaud Lambiotte},
  {and} \bibinfo{person}{Etienne Lefebvre}.} \bibinfo{year}{2008}\natexlab{}.
\newblock \showarticletitle{Fast unfolding of communities in large networks}.
\newblock \bibinfo{journal}{\emph{Journal of statistical mechanics: theory and
  experiment}} \bibinfo{volume}{2008}, \bibinfo{number}{10}
  (\bibinfo{year}{2008}), \bibinfo{pages}{P10008}.
\newblock


\bibitem[\protect\citeauthoryear{Chen, Sanghavi, and Xu}{Chen
  et~al\mbox{.}}{2014}]%
        {ipca}
\bibfield{author}{\bibinfo{person}{Yudong Chen}, \bibinfo{person}{Sujay
  Sanghavi}, {and} \bibinfo{person}{Huan Xu}.} \bibinfo{year}{2014}\natexlab{}.
\newblock \showarticletitle{Improved graph clustering}.
\newblock \bibinfo{journal}{\emph{IEEE Transactions on Information Theory}}
  \bibinfo{volume}{60}, \bibinfo{number}{10} (\bibinfo{year}{2014}),
  \bibinfo{pages}{6440--6455}.
\newblock


\bibitem[\protect\citeauthoryear{Diestel}{Diestel}{2006}]%
        {diestel2006graph}
\bibfield{author}{\bibinfo{person}{R. Diestel}.}
  \bibinfo{year}{2006}\natexlab{}.
\newblock \bibinfo{booktitle}{\emph{Graph Theory}}.
\newblock \bibinfo{publisher}{Springer}.
\newblock
\showISBNx{9783540261834}
\showLCCN{99057468}
\urldef\tempurl%
\url{https://books.google.co.in/books?id=aR2TMYQr2CMC}
\showURL{%
\tempurl}


\bibitem[\protect\citeauthoryear{Douce, Livingstone, and Orwell}{Douce
  et~al\mbox{.}}{2005}]%
        {10.1145/1163405.1163409}
\bibfield{author}{\bibinfo{person}{Christopher Douce}, \bibinfo{person}{David
  Livingstone}, {and} \bibinfo{person}{James Orwell}.}
  \bibinfo{year}{2005}\natexlab{}.
\newblock \showarticletitle{Automatic Test-Based Assessment of Programming: A
  Review}.
\newblock  \bibinfo{volume}{5}, \bibinfo{number}{3} (\bibinfo{year}{2005}).
\newblock
\showISSN{1531-4278}
\urldef\tempurl%
\url{https://doi.org/10.1145/1163405.1163409}
\showDOI{\tempurl}


\bibitem[\protect\citeauthoryear{HackerRank}{HackerRank}{[n. d.]}]%
        {hackerrank}
\bibfield{author}{\bibinfo{person}{HackerRank}.} \bibinfo{year}{[n.
  d.]}\natexlab{}.
\newblock \bibinfo{title}{HackerRank}.
\newblock   (\bibinfo{date}{April} \bibinfo{year}{[n. d.]}).
\newblock
\urldef\tempurl%
\url{http://hackerrank.com}
\showURL{%
Retrieved 2019-04-4 from \tempurl}


\bibitem[\protect\citeauthoryear{Jackson}{Jackson}{1996}]%
        {10.1016/S0360-1315(96)00025-5}
\bibfield{author}{\bibinfo{person}{David Jackson}.}
  \bibinfo{year}{1996}\natexlab{}.
\newblock \showarticletitle{A Software System for Grading Student Computer
  Programs}.
\newblock \bibinfo{journal}{\emph{Comput. Educ.}} \bibinfo{volume}{27},
  \bibinfo{number}{3–4} (\bibinfo{date}{Dec.} \bibinfo{year}{1996}),
  \bibinfo{pages}{171–180}.
\newblock
\showISSN{0360-1315}
\urldef\tempurl%
\url{https://doi.org/10.1016/S0360-1315(96)00025-5}
\showDOI{\tempurl}


\bibitem[\protect\citeauthoryear{Joy, Griffiths, and Boyatt}{Joy
  et~al\mbox{.}}{2005}]%
        {10.1145/1163405.1163407}
\bibfield{author}{\bibinfo{person}{Mike Joy}, \bibinfo{person}{Nathan
  Griffiths}, {and} \bibinfo{person}{Russell Boyatt}.}
  \bibinfo{year}{2005}\natexlab{}.
\newblock \showarticletitle{The Boss Online Submission and Assessment System}.
\newblock  \bibinfo{volume}{5}, \bibinfo{number}{3} (\bibinfo{year}{2005}).
\newblock
\showISSN{1531-4278}
\urldef\tempurl%
\url{https://doi.org/10.1145/1163405.1163407}
\showDOI{\tempurl}


\bibitem[\protect\citeauthoryear{Liu, Chen, Han, and Yu}{Liu
  et~al\mbox{.}}{2006}]%
        {GPLAG}
\bibfield{author}{\bibinfo{person}{Chao Liu}, \bibinfo{person}{Chen Chen},
  \bibinfo{person}{Jiawei Han}, {and} \bibinfo{person}{Philip~S. Yu}.}
  \bibinfo{year}{2006}\natexlab{}.
\newblock \showarticletitle{GPLAG: Detection of Software Plagiarism by Program
  Dependence Graph Analysis}. In \bibinfo{booktitle}{\emph{Proceedings of the
  12th ACM SIGKDD International Conference on Knowledge Discovery and Data
  Mining}} \emph{(\bibinfo{series}{KDD '06})}. \bibinfo{publisher}{ACM},
  \bibinfo{address}{USA}, \bibinfo{pages}{872--881}.
\newblock
\showISBNx{1-59593-339-5}
\urldef\tempurl%
\url{https://doi.org/10.1145/1150402.1150522}
\showDOI{\tempurl}


\bibitem[\protect\citeauthoryear{Luxburg}{Luxburg}{2007}]%
        {10.1007/s11222-007-9033-z}
\bibfield{author}{\bibinfo{person}{Ulrike Luxburg}.}
  \bibinfo{year}{2007}\natexlab{}.
\newblock \showarticletitle{A Tutorial on Spectral Clustering}.
\newblock  \bibinfo{volume}{17}, \bibinfo{number}{4} (\bibinfo{year}{2007}).
\newblock
\showISSN{0960-3174}
\urldef\tempurl%
\url{https://doi.org/10.1007/s11222-007-9033-z}
\showDOI{\tempurl}


\bibitem[\protect\citeauthoryear{MikeMirzayanov}{MikeMirzayanov}{[n. d.]}]%
        {CodeForces}
\bibfield{author}{\bibinfo{person}{MikeMirzayanov}.} \bibinfo{year}{[n.
  d.]}\natexlab{}.
\newblock \bibinfo{title}{CodeForces}.
\newblock   (\bibinfo{date}{June} \bibinfo{year}{[n. d.]}).
\newblock
\urldef\tempurl%
\url{http://codeforces.com}
\showURL{%
Retrieved 2018-06-4 from \tempurl}


\bibitem[\protect\citeauthoryear{Naud\'{e}, Greyling, and Vogts}{Naud\'{e}
  et~al\mbox{.}}{2010}]%
        {10.1016/j.compedu.2009.09.005}
\bibfield{author}{\bibinfo{person}{Kevin~A. Naud\'{e}},
  \bibinfo{person}{Jean~H. Greyling}, {and} \bibinfo{person}{Dieter Vogts}.}
  \bibinfo{year}{2010}\natexlab{}.
\newblock \showarticletitle{Marking Student Programs Using Graph Similarity}.
\newblock  \bibinfo{volume}{54}, \bibinfo{number}{2} (\bibinfo{year}{2010}).
\newblock
\showISSN{0360-1315}
\urldef\tempurl%
\url{https://doi.org/10.1016/j.compedu.2009.09.005}
\showDOI{\tempurl}


\bibitem[\protect\citeauthoryear{Novak, Joy, and Kermek}{Novak
  et~al\mbox{.}}{2019}]%
        {10.1145/3313290}
\bibfield{author}{\bibinfo{person}{Matija Novak}, \bibinfo{person}{Mike Joy},
  {and} \bibinfo{person}{Dragutin Kermek}.} \bibinfo{year}{2019}\natexlab{}.
\newblock \showarticletitle{Source-Code Similarity Detection and Detection
  Tools Used in Academia: A Systematic Review}.
\newblock  \bibinfo{volume}{19}, \bibinfo{number}{3} (\bibinfo{year}{2019}).
\newblock
\urldef\tempurl%
\url{https://doi.org/10.1145/3313290}
\showDOI{\tempurl}


\bibitem[\protect\citeauthoryear{Prechelt and Malpohl}{Prechelt and
  Malpohl}{2003}]%
        {jplag}
\bibfield{author}{\bibinfo{person}{Lutz Prechelt} {and} \bibinfo{person}{Guido
  Malpohl}.} \bibinfo{year}{2003}\natexlab{}.
\newblock \showarticletitle{Finding Plagiarisms among a Set of Programs with
  JPlag}.
\newblock \bibinfo{journal}{\emph{Journal of Universal Computer Science}}
  \bibinfo{volume}{8} (\bibinfo{date}{03} \bibinfo{year}{2003}).
\newblock


\bibitem[\protect\citeauthoryear{Ribeiro and Guerreiro}{Ribeiro and
  Guerreiro}{2009}]%
        {ribeiro2009improving}
\bibfield{author}{\bibinfo{person}{Pedro Ribeiro} {and} \bibinfo{person}{Pedro
  Guerreiro}.} \bibinfo{year}{2009}\natexlab{}.
\newblock \showarticletitle{Improving the automatic evaluation of problem
  solutions in programming contests}.
\newblock \bibinfo{journal}{\emph{Olympiads in Informatics}}
  \bibinfo{volume}{3} (\bibinfo{year}{2009}), \bibinfo{pages}{132--143}.
\newblock


\bibitem[\protect\citeauthoryear{Roy, Cordy, and Koschke}{Roy
  et~al\mbox{.}}{2009}]%
        {10.1016/j.scico.2009.02.007}
\bibfield{author}{\bibinfo{person}{Chanchal~K. Roy}, \bibinfo{person}{James~R.
  Cordy}, {and} \bibinfo{person}{Rainer Koschke}.}
  \bibinfo{year}{2009}\natexlab{}.
\newblock \showarticletitle{Comparison and Evaluation of Code Clone Detection
  Techniques and Tools: A Qualitative Approach}.
\newblock  \bibinfo{volume}{74}, \bibinfo{number}{7} (\bibinfo{year}{2009}).
\newblock
\showISSN{0167-6423}
\urldef\tempurl%
\url{https://doi.org/10.1016/j.scico.2009.02.007}
\showDOI{\tempurl}


\bibitem[\protect\citeauthoryear{Schleimer, Wilkerson, and Aiken}{Schleimer
  et~al\mbox{.}}{2003}]%
        {Schleimer:2003:WLA:872757.872770}
\bibfield{author}{\bibinfo{person}{Saul Schleimer}, \bibinfo{person}{Daniel~S.
  Wilkerson}, {and} \bibinfo{person}{Alex Aiken}.}
  \bibinfo{year}{2003}\natexlab{}.
\newblock \showarticletitle{Winnowing: Local Algorithms for Document
  Fingerprinting}. In \bibinfo{booktitle}{\emph{Proceedings of the 2003 ACM
  SIGMOD International Conference on Management of Data}}
  \emph{(\bibinfo{series}{SIGMOD '03})}. \bibinfo{publisher}{ACM},
  \bibinfo{address}{New York, NY, USA}, \bibinfo{pages}{76--85}.
\newblock
\showISBNx{1-58113-634-X}
\urldef\tempurl%
\url{https://doi.org/10.1145/872757.872770}
\showDOI{\tempurl}


\bibitem[\protect\citeauthoryear{Shivam, Goswami, Baths, and
  Bandyopadhyay}{Shivam et~al\mbox{.}}{2019}]%
        {DBLP:conf/icsoft/ShivamGBB19}
\bibfield{author}{\bibinfo{person}{Shivam}, \bibinfo{person}{Nilanjana
  Goswami}, \bibinfo{person}{Veeky Baths}, {and} \bibinfo{person}{Soumyadip
  Bandyopadhyay}.} \bibinfo{year}{2019}\natexlab{}.
\newblock \showarticletitle{{AES:} Automated Evaluation Systems for Computer
  Programing Course}. In \bibinfo{booktitle}{\emph{Proceedings of the 14th
  International Conference on Software Technologies, {ICSOFT} 2019, Prague,
  Czech Republic, July 26-28, 2019}}, \bibfield{editor}{\bibinfo{person}{Marten
  van Sinderen} {and} \bibinfo{person}{Leszek~A. Maciaszek}} (Eds.).
  \bibinfo{publisher}{SciTePress}, \bibinfo{pages}{508--513}.
\newblock
\urldef\tempurl%
\url{https://doi.org/10.5220/0007951205080513}
\showDOI{\tempurl}


\bibitem[\protect\citeauthoryear{Singh, Srikant, and Aggarwal}{Singh
  et~al\mbox{.}}{2016}]%
        {quesindep}
\bibfield{author}{\bibinfo{person}{Gursimran Singh}, \bibinfo{person}{Shashank
  Srikant}, {and} \bibinfo{person}{Varun Aggarwal}.}
  \bibinfo{year}{2016}\natexlab{}.
\newblock \showarticletitle{Question Independent Grading using Machine
  Learning: The Case of Computer Program Grading}. In
  \bibinfo{booktitle}{\emph{Proceedings of the 22nd ACM SIGKDD International
  Conference on Knowledge Discovery and Data Mining}}.
  \bibinfo{pages}{263--272}.
\newblock
\urldef\tempurl%
\url{https://doi.org/10.1145/2939672.2939696}
\showDOI{\tempurl}


\bibitem[\protect\citeauthoryear{Singh, Gulwani, and Solar-Lezama}{Singh
  et~al\mbox{.}}{2013}]%
        {10.1145/2499370.2462195}
\bibfield{author}{\bibinfo{person}{Rishabh Singh}, \bibinfo{person}{Sumit
  Gulwani}, {and} \bibinfo{person}{Armando Solar-Lezama}.}
  \bibinfo{year}{2013}\natexlab{}.
\newblock \showarticletitle{Automated Feedback Generation for Introductory
  Programming Assignments}.
\newblock  \bibinfo{volume}{48}, \bibinfo{number}{6} (\bibinfo{year}{2013}).
\newblock
\showISSN{0362-1340}
\urldef\tempurl%
\url{https://doi.org/10.1145/2499370.2462195}
\showDOI{\tempurl}


\bibitem[\protect\citeauthoryear{Srikant and Aggarwal}{Srikant and
  Aggarwal}{2014}]%
        {srikant}
\bibfield{author}{\bibinfo{person}{Shashank Srikant} {and}
  \bibinfo{person}{Varun Aggarwal}.} \bibinfo{year}{2014}\natexlab{}.
\newblock \showarticletitle{A system to grade computer programming skills using
  machine learning}.
\newblock \bibinfo{journal}{\emph{Proceedings of the ACM SIGKDD International
  Conference on Knowledge Discovery and Data Mining}} (\bibinfo{year}{2014}).
\newblock
\showISBNx{978-1-4503-2956-9}
\urldef\tempurl%
\url{https://doi.org/10.1145/2623330.2623377}
\showDOI{\tempurl}


\bibitem[\protect\citeauthoryear{von Matt}{von Matt}{1994}]%
        {10.1145/182107.182101}
\bibfield{author}{\bibinfo{person}{urs von Matt}.}
  \bibinfo{year}{1994}\natexlab{}.
\newblock \showarticletitle{Kassandra: The Automatic Grading System}.
\newblock  \bibinfo{volume}{22}, \bibinfo{number}{1} (\bibinfo{year}{1994}).
\newblock
\showISSN{0163-5735}
\urldef\tempurl%
\url{https://doi.org/10.1145/182107.182101}
\showDOI{\tempurl}


\bibitem[\protect\citeauthoryear{Wang, Su, Wang, and Ma}{Wang
  et~al\mbox{.}}{2007}]%
        {10.1016/j.infsof.2006.03.001}
\bibfield{author}{\bibinfo{person}{Tiantian Wang}, \bibinfo{person}{Xiaohong
  Su}, \bibinfo{person}{Yuying Wang}, {and} \bibinfo{person}{Peijun Ma}.}
  \bibinfo{year}{2007}\natexlab{}.
\newblock \showarticletitle{Semantic Similarity-Based Grading of Student
  Programs}.
\newblock  \bibinfo{volume}{49}, \bibinfo{number}{2} (\bibinfo{year}{2007}).
\newblock
\showISSN{0950-5849}
\urldef\tempurl%
\url{https://doi.org/10.1016/j.infsof.2006.03.001}
\showDOI{\tempurl}


\bibitem[\protect\citeauthoryear{Wick, Stevenson, and Wagner}{Wick
  et~al\mbox{.}}{2005}]%
        {10.1145/1047124.1047427}
\bibfield{author}{\bibinfo{person}{Michael Wick}, \bibinfo{person}{Daniel
  Stevenson}, {and} \bibinfo{person}{Paul Wagner}.}
  \bibinfo{year}{2005}\natexlab{}.
\newblock \showarticletitle{Using Testing and JUnit across the Curriculum}.
\newblock  \bibinfo{volume}{37}, \bibinfo{number}{1} (\bibinfo{year}{2005}).
\newblock
\showISSN{0097-8418}
\urldef\tempurl%
\url{https://doi.org/10.1145/1047124.1047427}
\showDOI{\tempurl}


\end{thebibliography}
\end{document}